\documentclass[print, sort&compress]{elsarticle}

\usepackage{lineno}

\usepackage{graphicx}
\usepackage{dcolumn}
\usepackage{bm}
\usepackage{upgreek}
\usepackage{hyperref}
\usepackage{amsmath}
\usepackage{multirow}
\usepackage[margin=1.3in]{geometry}
\let\oldequation\equation
\let\oldendequation\endequation

\renewenvironment{equation}
  {\linenomathNonumbers\oldequation}
  {\oldendequation\endlinenomath}

\usepackage{multicol} 

\usepackage{nomencl} 

\makenomenclature

\usepackage{soul,color,xcolor}
\sethlcolor{white}
\soulregister{\cite}7
\soulregister{\citep}7
\soulregister{\citet}7
\soulregister{\ref}7
\soulregister{\pageref}7

\usepackage{etoolbox}
\renewcommand\nomgroup[1]{%
  \item[\bfseries
  \ifstrequal{#1}{G}{Greek symbols}{%
  \ifstrequal{#1}{S}{Subscripts}{}}%
]}

\journal{Chemical Engineering Science}

\usepackage{caption}

\begin{document}

\begin{frontmatter}

\title{Two-phase flow in porous metal foam flow fields of PEM fuel cells}

\author[TJU]{Xingxiao Tao}
\author[TJU]{Kai Sun}
\author[TJU,LU]{Rui Chen}
\author[TJU]{Mengshan Suo}
\author[TJU]{Huaiyu Liu}
\author[TJU,PES]{Zhizhao Che\corref{cor1}}
\ead{chezhizhao@tju.edu.cn}
\author[TJU,PES]{Tianyou Wang\corref{cor1}}
\ead{wangtianyou@tju.edu.cn}

\cortext[cor1]{Corresponding authors.}

\address[TJU]{State Key Laboratory of Engines, Tianjin University, Tianjin, 300350, China.}
\address[LU]{Department of Aeronautical and Automotive Engineering, Loughborough University, Loughborough LE11 3TU, United Kingdom}
\address[PES]{National Industry-Education Platform of Energy Storage, Tianjin University, Tianjin, 300350, China}

\begin{abstract}
Porous metal foam (PMF) flow field is a potential option for proton exchange membrane fuel cells (PEMFCs) due to its excellent capabilities in gas distribution and water drainage. However, the gas-liquid two-phase flow in the PMF flow field on the pore scale is still unclear. In this study, we investigate the gas-liquid two-phase flow in the PMF flow field. Film, plug, and ligament flows are found in the hydrophilic PMF flow field, while slug and droplet flows are found in the hydrophobic PMF flow field. The results suggest that optimizing the pore size, increasing the metal foam surface hydrophobicity, and optimizing the operating condition are helpful for the water management of the PMF flow field. The frequency analysis of the pressure drop also shows that the dominant frequency can be used as an indicator to analyze the transition between different flow patterns.
\end{abstract}

\begin{keyword}
\texttt {
Porous metal foam \sep
Fuel cell flow field \sep
Two-phase flow \sep
Pressure drop \sep
Proton exchange membrane fuel cell
}
\end{keyword}

\end{frontmatter}

\def \scaleSize {0.8}
\def \scaleSiz2 {0.6}

\section{Introduction}\label{sec1}
Water management is a crucial challenge for proton exchange membrane fuel cells (PEMFCs) before their successful commercialization \cite{Tao2023DesigningGradedFuelCell, Tao2023DesignofPorousFlowField, Tao2018PressureDropRegulationStrategy, Tao2021DesigningNextGenerationofPEMFC}. As a pivotal component of PEMFCs, the flow field plays a crucial role in reaction gas transport and product water drainage. An ideal flow field should aim for that reaction gas reaches the catalytic active reaction interface evenly without excessive pressure drop, and meanwhile can accelerate the removal of water to avoid flooding \cite{Sauermoser2020}. Therefore, it is significant to optimize the flow field structure and investigate the two-phase flow process inside the flow field to enhance the overall PEMFCs performance \cite{Tao2020BioInspiredWaveLikeFlowChannel, Tao2020MultiComponentMultiPhaseFlow, Tao2011WaterTransport, Tao2018FlowFieldGeometryDesigns, Tao2018BiPorousLayerFlowField, Tao2021Two-phaseFlowThroughOpen-cellMetalFoam, Tao2022Porous-RibFlowField, Tao2021FlowPatternOnFoamStructures, Tao2018MultiPhaseCFDModel, Tao20183DFineMeshFlowField}.

The ``channel-rib'' structure has been widely adopted in conventional PEMFC flow fields. However, it results in the reactants' mal-distribution under the ``channel'' and the ``rib'', which includes water flooding under the rib. The distribution of water and the characteristics of gas-liquid two-phase flow in the conventional channels of PEMFCs were of great concern in the past few years \cite{Tao2017FlowFieldOrientation, Tao2014TwoPhaseFlowVisualization, Tao2021DropletDynamics, Tao2023Large-sizedPEMFCFlowDistribution, Tao2009ChannelFloodingVisualization, Tao2009TwoPhaseFlow, Tao2019PorousFlowFieldExperimentalStudy, Tao2015WaterSlugMotion, Tao2022Two-phaseFlowinParallelChannels}. Banerjee et al.\ \cite{Tao2014TwoPhaseFlowVisualization} proposed a new method to evaluate the water in the cathode channel quantitatively based on area coverage ratio (ACR), which revealed the two-phase flow pattern under different working conditions. Hussaini et al.\ \cite{Tao2009ChannelFloodingVisualization} analyzed the water flooding in several parallel channels on the cathode by in-situ visualization. They determined four flow patterns for a range of conventional automotive working conditions and constructed the flow pattern map in an operating PEMFC. Lu et al.\ \cite{Tao2009TwoPhaseFlow} observed the mist flow besides the slug and the film flow in the parallel channel via ex-situ visualization. Their result demonstrated that the film flow is the beneficial flow pattern for water removal, the slug flow can lead to serious flow mal-distribution and large pressure fluctuations, and the mist flow may cause high parasitic power loss because of the higher airflow rate. Ye et al.\ \cite{Tao2015WaterSlugMotion} correlated the images of the formation and motion of slugs in microfluidic systems with the pressure/flow results obtained by synchronized measurements. They found that the gas flow path inside the channel is determined by many elements, including the gas diffusion layer (GDL) permeability and the slug volume.

Recently, porous metal foam (PMF) flow field \cite{Tao2010PorousMediaTwoPhaseFlow, Tao2017ColdStartPerformance, Tao2019PhaseTransitionMechanism, Tao2018GrapheneFoamFlowField, Tao2022AnOpen-cellFoamsModelling, Tao2010PorousCopperFiber, Tao2004MetalFoamDesign, Tao2017OpenPoreCellularFoam} has been proposed to provide a new route to improve and optimize the PEMFC flow field due to its excellent gas distribution and water drainage capabilities, good thermal and electrical conductivity \cite{Tao2016MetalFoamCoolingFlowField, Tao2022MetalFoamsonSubcooledFlowBoiling}, not only in fuel cells but also in other applications such as heat exchangers, energy absorbers and filters. Li et al.\ \cite{Tao2019FlowBoiling} found that the flow patterns during boiling heat transfer of refrigerant R141b inside the parallel channel filled with copper foam are similar to that in horizontal macro channels. But the occurrence time and boundary conditions between different flow patterns are different. Shin et al.\ \cite{Tao2018FoamSizeEffect}, by using the combination of metal foams of different pore sizes as the flow field of PEMFCs, reduced the fluctuation of output current and improved the performance and stability of the fuel cell. Kim et al.\ \cite{Tao2017FoaminPEMFCsStack} used metal foam as the flow field of a small fuel cell stack (500 W) and confirmed the feasibility of metal foam flow field in fuel cell stacks. Bao et al.\ \cite{Tao2019MetalFoamFlowField} numerically analyzed the detailed two-phase flow in the pore-scale of the metal foam (MF) flow field with reconstructed the full three-dimensional morphology of MF. They found that MF can reduce the water coverage at the GDL/channel interface and enhance gas convection. Further, they reconstructed the MF flow field model via X-ray morphological analysis of nickel foam samples and used it to simulate the gas transport mode \cite{Tao2020TwoPhaseMassTransport} and the detachment and dispersion of droplets \cite{Tao2020LiquidDispersion} in the MF flow field. Their simulation proved that the MF flow field has lower gas permeability than the conventional channel. Their results demonstrate that the MF flow field is fit for the operating conditions of high current density. Chen et al.\ \cite{Tao2010PorousMediaTwoPhaseFlow} studied the influence of inserting porous carbon foam on the uniformity of two-phase flow distribution in conventional parallel channels. Tabe et al.\ \cite{Tao2013PorousFlowField} investigated the water behavior in the PMF flow field in an operating PEMFC from the perspective of the cross-section. They found that porous cell is beneficial to the condensed water drained out from the GDL surface. Besides, the overall PEMFCs performance especially under the high current density will be seriously hampered by the surface wettability of porous materials. Fly et al.\ \cite{Tao2018FlowDistribution} used ex-situ optical testing and X-ray Computed Tomography (CT) to evaluate the flow rate and pressure difference distribution of five different PMF flow fields quantitatively and proposed a mixed-type flow field design. Their results \cite{Tao2019TwoPhaseNumericalModel} further explained the influence of the foam structure and air velocity on the water from GDL into the PMF flow field.

It can be found that most of the existing studies of two-phase flow in PEMFCs concentrate on the conventional ``channel-rib'' channel. The few studies of the PMF flow field only consider the particular gas-liquid velocity and foam size \cite{Tao2010PorousMediaTwoPhaseFlow}. The accurate identification of two-phase flow patterns in the PMF flow field in different conditions is vital for the design of the PEMFC flow field. In general, the two-phase flow pattern in PEMFCs is influenced by many factors, which include the gas and liquid flow rates, the surface wettability, and the channel geometry \cite{Tao2021TwoPhaseFlowMachineLearningApplications}. Hence, the main objective of this study is to investigate the gas-liquid two-phase flow in the PMF flow field via optical visualization and pressure drop measurement under PEMFC flow conditions and structure parameters, aiming to optimize the foam pore size, surface wettability, and operating conditions, and to provide guidance for the PMF flow field design. The rest of this paper is organized as follows. Experimental details are discussed in Section \ref{sec2}. The results are discussed in Section \ref{sec3}. The flow pattern and the flow resistance are analyzed, and the effects of the water and air flow rates, the pore size, and the surface wettability are considered. Finally, conclusions are drawn in Section \ref{sec4}.

\section{Experimental setup}\label{sec2}
\begin{figure}
  \centering
  \includegraphics[scale=0.5]{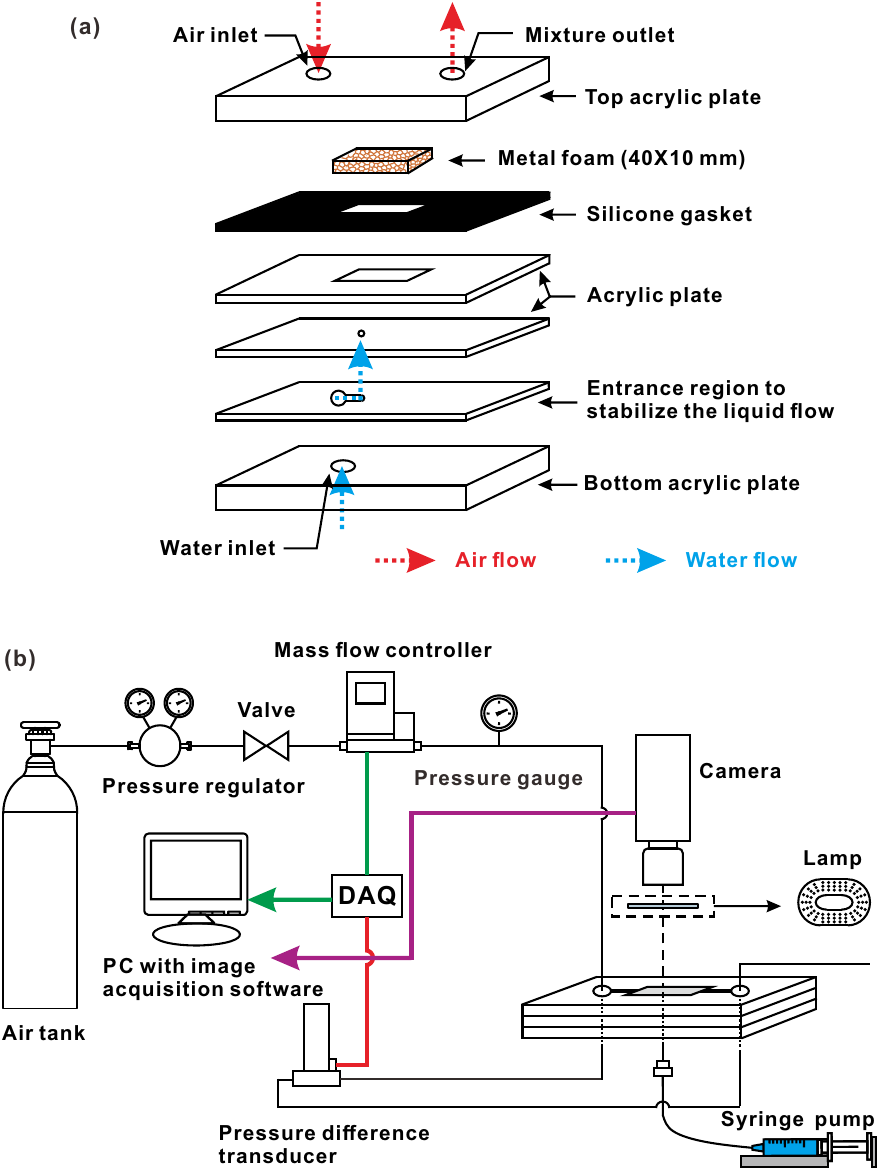}
  \caption{(a) Schematic diagram of the transparent assembly of the PMF flow field structure. (b) Schematic diagram of the test loop.}\label{fig:01}
\end{figure}

The experimental setup is schematically shown in Fig.\ \ref{fig:01}a. A test assembly was designed to represent the PMF flow field structure of PEMFCs, and it consists of a sample of copper metal foam sandwiched between silicone gaskets and transparent acrylic plates. The metal foam sample is 40 mm in length in the flow direction, 10 mm in width, and 3 mm in thickness. A hole with a radius of 0.5 mm was drilled on the acrylic plate at the bottom of the metal foam, which was used as a water injection port to simplify the process of water introduction from the gas diffusion layer (GDL) to the flow field. The directions of the airflow inlet and the liquid flow inlet were perpendicular to each other. The two-phase flow in the PMF flow field was visualized through the transparent acrylic plate at the top.

The experimental system is illustrated in Fig.\ \ref{fig:01}b. The system simultaneously measured the two-phase (air and water) flow rate and the pressure drop between the inlet and the outlet of the PMF flow field. A CMOS camera (FLIR, Grasshopper GS3-U3-123S6C) with a macro lens (Tokina MACRO 100mm F2.8 D) was used to capture the top-view image of the two-phase flow process in the PMF flow field, with the illumination of a macro photographic lamp. The camera was set at a frame rate of 50 frames per second (fps), with spatial resolutions of 8.06 $\upmu$m pixel$^{-1}$ for the top-view recording. A syringe pump (Harvard Apparatus Pump 11 Pico Plus Elite) was used to control the liquid water supplied at a flow rate ranging from 0.002 to 0.013 m s$^{-1}$ at the inlet of foam, which covered all the actual working conditions of PEMFCs from drying to flooding. To better distinguish water from the air in the experimental images, blue dye (Mkm/Ameresco Coomassie brilliant blue G-250) was added to the water in some of the experiments. The airflow rate was regulated by a gas mass flow controller (Alicat MC-D-DB9M-ACDS for low flow rates, KONXIN YJ-700CT for high flow rates) ranging from 1 to 8 L min$^{-1}$, which was selected according to the typical operating conditions of PEMFCs \cite{Tao2009Two-phaseFlowParallelChannels}. The measurement accuracy of the syringe pump and gas mass flow controller was $\pm$0.35\% and $\pm$0.1\% of the full scale, respectively. The data collection system includes a pulse signal generator (Quantum 9214), a differential pressure transducer (ZHONG HUAN TIG TDS4433-1CA00-1AA6A01), and a data acquisition instrument (National Instruments, NI USB-6361). The pressure drop was measured by the pressure difference transducer with a maximum calibration ranging from 0 to 20 kPa and an accuracy of $\pm$0.2\% of full scale. The camera and the pressure difference transducer were triggered synchronously by the signal generator. Measurement data from the transducers were real-time displayed and recorded. The pressure drop data and images were processed by a customized Matlab program and ImageJ. All images were taken after the system was stable. All experiments were performed at room temperature (23--27 $^\circ$C).

\begin{figure}
  \centering
  \includegraphics[scale=0.7]{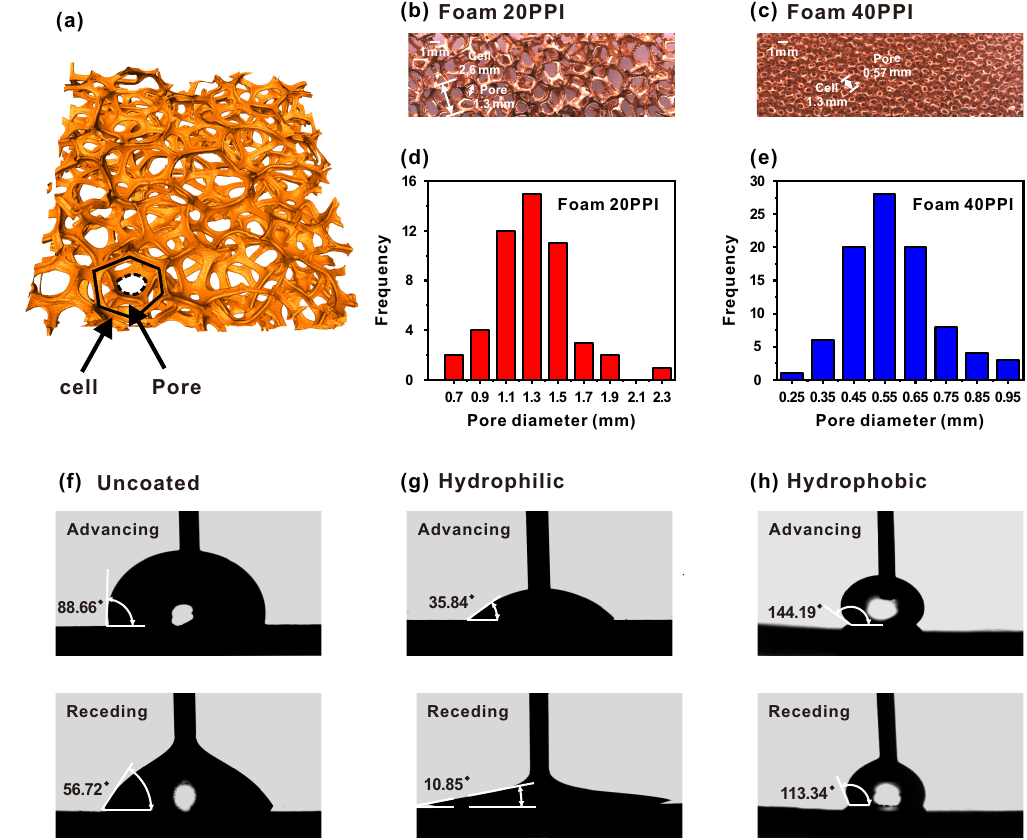}
  \caption{Metal foam structure used in the experiment: (a) Metal foam structure obtained by
X-ray tomography. (b-c) Images of pore morphology for foams of 20 and 40 PPI. (d-e) Pore size distribution for foams of 20 and 40 PPI. (f-h) Advancing and receding contact angles of liquid water on copper surfaces with different wettabilities. The substrates beneath the droplets are smooth copper surfaces, and the contact angles in the images are the intrinsic contact angles. }\label{fig:02}
\end{figure}

To characterize the structure of the PMF flow field, a metal foam structure was reconstructed by X-ray tomography with a microCT (nanoVoxel-3000, Sanying Precision Instruments) with a spatial resolution of 2.965 $\upmu$m, as shown in Fig.\ \ref{fig:02}a. Considering the difficulty of optical visualization caused by the existence of a large number of pore structures in the metal foam and the blocking effect between different foam ligaments, two types of metal foam samples were used in the experiments, as shown in Figs.\ \ref{fig:02}b and \ref{fig:02}c, namely copper foams 20 PPI and 40 PPI. The pore size distribution of the metal foams is shown in Figs.\ \ref{fig:02}(d-e). The average pore sizes are 1.31 and 0.57 mm for the foams 20 PPI and 40 PPI, respectively. The porosity of the foam is 95\%--98\%, and the area density is 1800 g m$^{-2}$. It is worth noting that the thickness of the metal foam selected here is 3 mm, which is mainly to capture more flow details in the PMF flow field clearly.

To study the effect of the surface wettability on the two-phase flow in the PMF flow field, in some experiments, the surface of metal foam with a pore density of 20 PPI (pores per inch) was modified by chemical etching methods \cite{Tao2009CopperSurfaceWettabilityFabrication, Tao2017DrainageCharacteristic}. The hydrophilic copper foam was fabricated via alkali-assisted surface oxidation, and the hydrophobic copper foam was treated by soaking in n-dodecyl mercaptan ethanol solution after surface oxidation \cite{Tao2009CopperSurfaceWettabilityFabrication, Tao2017DrainageCharacteristic}. It should be pointed out that the metal foam before the chemical modification has the same initial porosity (95\%--98\%), and the effect of the chemical etching method on the porosity of the metal foam can be ignored. The porous materials' contact angles include the intrinsic contact angle determined by the material and the apparent contact angle affected by the geometric structure. In Figs.\ \ref{fig:02}(f-h), we use the intrinsic contact angle to characterize the surface wettability of different metal foams. Considering that the intrinsic contact angle is only determined by the characteristics of the material itself, we used copper plates with the same purity (99.99\%), carried out the same surface treatment, and then measured the advancing and receding contact angles of the copper plates before and after the surface modification through the method of droplet growing/shrinking, as shown in Figs.\ \ref{fig:02}(f-h).

\section{Results and discussion}\label{sec3}
\subsection{Two-phase flow patterns }\label{sec31}
\subsubsection{Typical flow patterns}\label{sec311}
\begin{figure}
  \centering
  \includegraphics[scale=0.6]{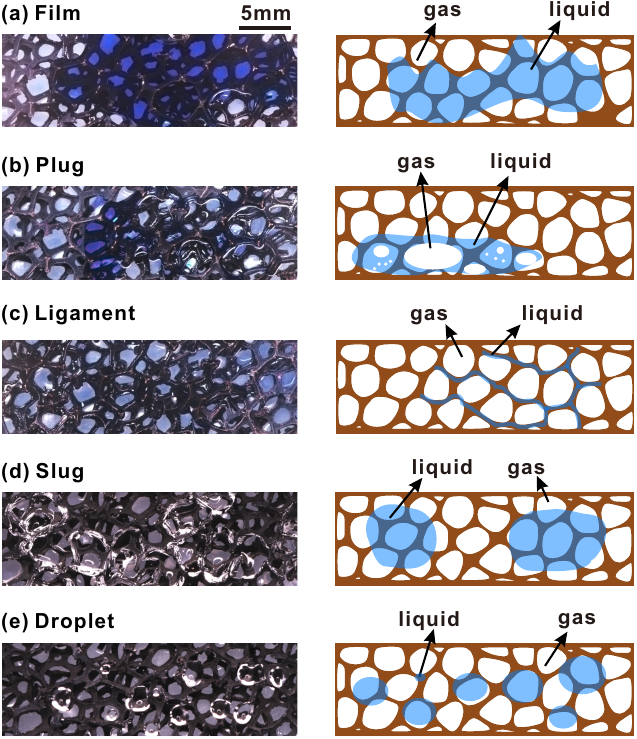}
  \caption{Flow patterns in the PMF flow field. (a) Film flow; (b) Plug flow; (c) Ligament flow; (d) Slug flow; (e) Droplet flow. The left column is experimental images, and the right column is schematic illustrations. In the illustrations, the brown, blue, and white colors indicate ligaments of the metal foam, liquid water, and gas, respectively. Images in Figs.\ \ref{fig:03}(a-c) were taken in the hydrophilic PMF flow field with blue dye to increase the contrast, and images in Figs.\ \ref{fig:03}(d-e) were taken in the hydrophobic PMF flow field without dye. The superficial gas velocity ($U_{sg}$) and the superficial liquid velocity ($U_{sl}$) are respectively: (a) $U_{sg}$ = 0.56 m s$^{-1}$, $U_{sl}$ = 0.0043 m s$^{-1}$; (b) $U_{sg}$ = 2.22 m s$^{-1}$, $U_{sl}$ = 0.0043 m s$^{-1}$; (c) $U_{sg}$ = 4.44 m s$^{-1}$, $U_{sl}$ = 0.0021 m s$^{-1}$; (d) $U_{sg}$ = 0.56 m s$^{-1}$, $U_{sl}$ = 0.0043 m s$^{-1}$; (e) $U_{sg}$ = 4.44 m s$^{-1}$, $U_{sl}$ = 0.0021 m s$^{-1}$. }\label{fig:03}
\end{figure}

Typical two-phase flow patterns in the conventional ``channel-rib'' channel of PEMFC \cite{Tao2010TwoPhaseFlowReview} include droplet flow, annular (or stratified) flow, and slug flow \cite{Tao2010PorousMediaTwoPhaseFlow, Tao2009ChannelFloodingVisualization, Tao2012MultiphaseFlow}. Our experiment shows that the PMF flow field can remarkably affect the flow pattern due to the complexity of metal ligaments and the randomness of pore structures. For PMF flow fields with different surface wettabilities, the flow pattern can be divided into five regimes: film flow, plug flow, ligament flow, slug flow, and droplet flow, as shown in Fig.\ \ref{fig:03}.

When the foam surface is \emph{hydrophilic}, liquid water may form a continuous film flow along the pores and intricate ligaments, where minute bubbles are dispersed in the liquid film. With the increase of the airflow rate, the number and the volume of minute bubbles gradually increase, forming discontinuous air plugs in the continuous liquid film. As the airflow rate further increases, the flow pattern evolves to ligament flow. The liquid phase spreads slowly along the solid wall of the ligament due to the strong capillarity on the hydrophilic surface, and the gas phase flows through the pore centers.

When the foam surface is \emph{hydrophobic}, the flow becomes slug flow or droplet flow. The slug flow is characterized by the volume of liquid continuously gradually filling multiple pores in sequence, which is remarkably different from the spreading process of the liquid film in multiple directions in the hydrophilic PMF flow field. The growth of a single droplet or the accumulation of multiple droplets can form the slug flow. The elongation and detachment of the slug are affected by the airflow rate and contact angle hysteresis \cite{Tao2008MicroscaleTwoPhaseFlow}. With the increase of airflow velocity, the slugs are broken into many droplets, which are dispersed in the pores randomly, forming droplet flow. Some droplets move forward along the airflow, while other droplets adhere to the weak-flow areas \cite{Tao2020TwoPhaseMassTransport} located near the small foam pores and the solid ligament intersections due to the high surface adhesion on the metal ligament. The above two flow patterns, slug flow and droplet flow, mainly occur in the hydrophobic PMF flow field, not in the hydrophilic PMF flow field.

\subsubsection{Effect of pore size}\label{sec312}
\begin{figure}
  \centering
  \includegraphics[scale=0.8]{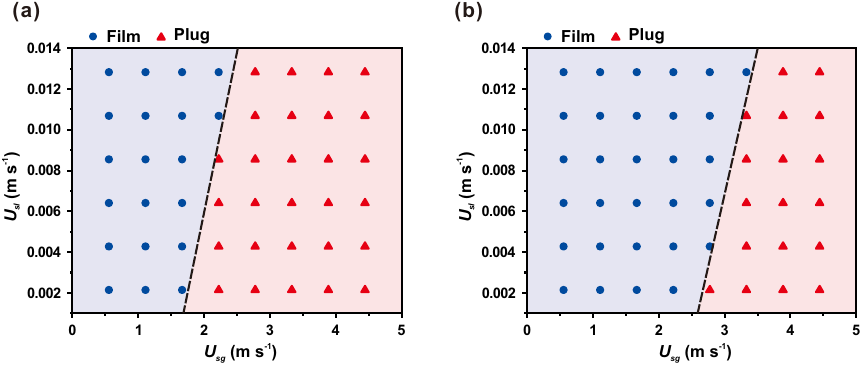}
  \caption{Flow pattern map plotted against the superficial gas velocity ($U_{sg}$) and the superficial liquid velocity ($U_{sl}$) in the PMF flow field with different pore sizes: (a) Foam 20 PPI, (b) Foam 40 PPI. The dashed lines are guidance for the eyes.}\label{fig:04}
\end{figure}

The pore size is a key parameter to the PMF flow field \cite{Tao2019OpenCellMetalFoamFlowField, Tao2019OptimizedFoamFlowField}. Here, flow pattern distribution maps are constructed to investigate the effect of metal foam pore size on two-phase flow and mass transfer. Different flow patterns are represented by a map using the gas and liquid superficial velocities as the axes. Two flow pattern maps are shown in Figs. \ref{fig:04}a and \ref{fig:04}b based on the visualization result of the foams of 20 PPI and 40 PPI, respectively. The critical air velocity for the transition from the film flow pattern to the plug flow pattern at the different liquid velocities is also provided.

As shown in Fig.\ \ref{fig:04}a and Fig.\ \ref{fig:04}b, a relatively higher superficial liquid water velocity requires a greater gas velocity for the transition from the film flow pattern to the plug flow pattern which is mainly because the increase of the hydrostatic pressure inhibits the growth and break of bubbles in the liquid film, thus increasing the required airflow rate to separate the continuous liquid film. A comparison between Figs.\ \ref{fig:04}a and \ref{fig:04}b shows that, as the pore size decreases, the film flow occupies a wider range of gas velocity. Smaller pore size may lead to an increase in the contact area between liquid water and foam ligament solid wall and, therefore, enhance the adhesion and spreading of the liquid film due to the surface tension. Besides, further reducing the foam pore size, the capillary pressure increases, thus promoting the spreading of the liquid film. This result suggests that in the range of parameters considered in this study, capillarity is dominant compared with the aerodynamic force resulting from the inlet airflow. If the pore size of the foam is further reduced, the critical airflow velocity required to remove the liquid is higher \cite{Tao2020LiquidDispersion} because of the severe liquid retention, which may cause the liquid water to be trapped in the pores and not spread easily.

\subsubsection{Effect of surface wettability}\label{sec313}
During the operation of PEMFCs, the flow field surface wettability significantly affects the water drained out from the GDL surface to the flow field. For the conventional ``channel-rib'' flow field, the hydrophilic flow channel can facilitate the detachment of water from the GDL surface \cite{Tao2013PorousFlowField}, while the hydrophobic flow channel is beneficial to drain water out of the flow channel. However, in the PMF flow field, owing to the complex and random solid foam ligament and pore structure, the mechanism of liquid water removal is completely different from that in the ``channel-rib''  flow field. The effect of surface wettability of ligaments of the PMF flow field on two-phase flow is studied in this section.

\begin{figure}
  \centering
  \includegraphics[scale=0.8]{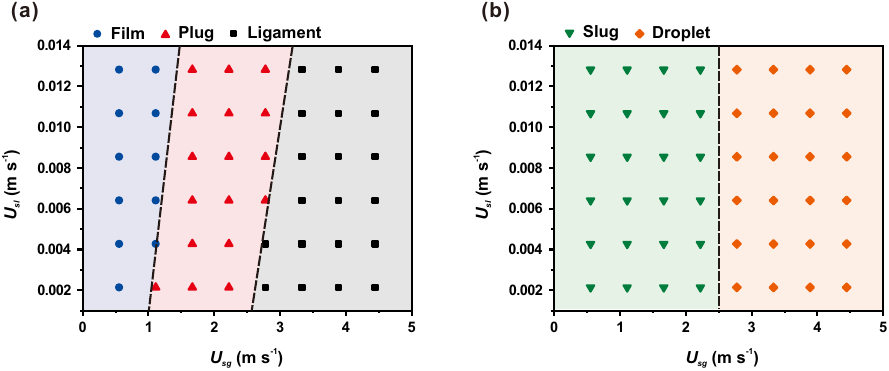}
  \caption{Flow pattern map plotted against the superficial gas velocity ($U_{sg}$) and the superficial liquid velocity ($U_{sl}$) in the PMF flow field with different surface wettabilities: (a) Hydrophilic, (b) Hydrophobic. The experiment was performed using Foam 20 PPI with surface modification as discussed in Figs.\ \ref{fig:02}(f-h). The dashed lines are guidance for the eyes.}\label{fig:05}
\end{figure}

The two-phase flow pattern maps in hydrophilic and hydrophobic PMF flow fields are shown in Figs.\ \ref{fig:05}a and \ref{fig:05}b, respectively. In the hydrophilic PMF flow field, three flow patterns were observed, including film, plug, and ligament flow, as shown in Fig.\ \ref{fig:05}a. Under a lower airflow rate, liquid droplets produced from the water injection port gradually grow and produce a large area of the liquid film after contacting the foam ligament, which is the typical film flow. The gas pathway in the hydrophilic PMF flow field is completely blocked by a large area of liquid film, and the residual amount of water is significantly higher than that in other flow patterns. Due to the large ligament surface adhesion force and air transport resistance, the film flow pattern leads to severe liquid retention in the PMF flow field. When further increasing the airflow, the growing gas plug in the liquid film cuts off a large piece of continuous liquid film, forming the plug flow, and the liquid water content in the PMF flow field decreases. Further increasing the airflow results in the transition to the ligament flow. In this condition, due to the strong ligament's surface adhesion, the water prefers to adhere to the ligament wall in the form of thin films, and there is only a small amount of liquid in the pores. Hence, in this condition, overcoming the liquid surface tension is the key to improving the drainage capacity of the hydrophilic PMF flow field.

In the hydrophobic PMF flow field, the flow pattern transition from the slug flow to the droplet flow is shown in Fig.\ \ref{fig:05}b. At a low airflow rate, the two-phase flow exhibits the slug flow pattern at all liquid velocities considered in this study. Liquid droplets are produced at the inlet and gradually fill part of the pores until forming large slugs. The volume of the slug increases with the continuous injection of liquid water, and gradually fills pores in sequence instead of filling several pores simultaneously. This is remarkably different from the spreading process of the liquid film in the hydrophilic PMF flow field, whereas the spreading process is in multiple directions due to the adhesion on the hydrophilic ligament surface. Hence, with the gradual filling process in the hydrophobic PMF flow field, which pore is to be filled next depends on the surface tension force and the air momentum in each pore \cite{Tao2019TwoPhaseNumericalModel}, which are both affected by the local pore structure. The slug is not discharged until the axial length of the slug extends to the PMF flow field outlet. Meanwhile, because of the weak adhesion of the ligament in the hydrophobic PMF flow field, liquid water is mainly retained in the pore center, increasing the blocking area to the airflow. Hence, this flow pattern reduces the gas drag force and increases the critical air superficial velocity required for liquid removal. At higher air velocities, the slug is deformed and broken into several randomly dispersed droplets due to the strong air aerodynamic force and foam ligament surface adhesion. As a consequence, the water coverage is reduced and the influence of liquid surface tension is weakened. Therefore, increasing the liquid velocity has little effect on the flow pattern distribution. The droplets tend to leave the initial location and be driven by the airflow when the aerodynamic force overcomes the adhesion force of the solid foam ligaments and the water surface tension in the pores. In this condition, the removal of water is mainly dependent on the airflow shear stress and the surface tension force considering the effect of contact angle hysteresis.

\subsection{Flow resistance analysis}\label{sec32}
\subsubsection{Two-phase friction multiplier}\label{sec321}
Because of the liquid water existence, the pressure drop of the two-phase flow in the flow field is significantly larger than that of the single-phase flow. The increased pressure drop owing to the product water accumulation in the flow field is the main reason for the flow mal-distribution of reaction gas in PEMFCs, which leads to the declination in the performance and the durability of the system \cite{Tao2009ChannelFloodingVisualization, Tao2017CathodePressureDropCalculate, Tao2009TwoPhaseFlow}. Previous studies have proved that the pressure drop ratio between two-phase flow and single-phase flow can be served as a characteristic of the degree of water accumulation in the PEMFC flow field. This ratio, defined as the two-phase friction multiplier ${\phi }_{g}^{2}$ \cite{Tao2009ChannelFloodingVisualization, Tao2009TwoPhaseFlow}, is defined as:
\begin{equation}\label{eq:1}
  {\phi }_{g}^{2}=\frac{\Delta {{P}_{2\phi }}}{\Delta {{P}_{g}}}
\end{equation}
where $\Delta {{P}_{2\phi }}$ and $\Delta {{P}_{g}}$ are the pressure drop with two-phase flow and with single-phase gas flow in the PMF flow field, respectively.

\begin{figure}
  \centering
  \includegraphics[scale=0.8]{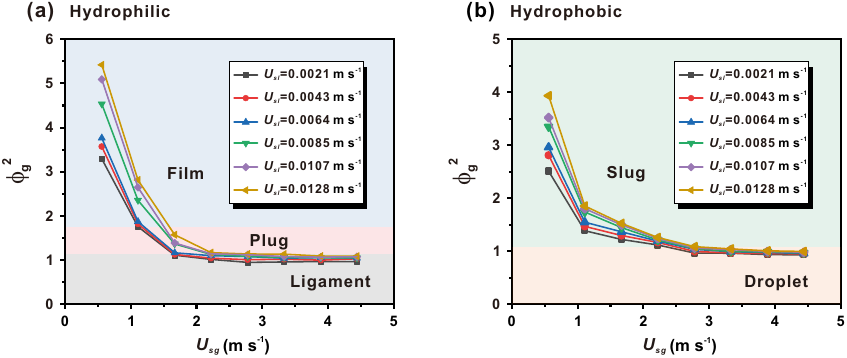}
  \caption{Two-phase friction multiplier ${\phi }_{g}^{2}$ versus the superficial gas velocity ($U_{sg}$) and the superficial liquid velocity ($U_{sl}$) for PMFs with different wettabilities: (a) Hydrophilic PMF; (b) Hydrophobic PMF. The different lines in the figures correspond to different superficial liquid velocities, and the different background colors indicate different flow patterns.}\label{fig:06}
\end{figure}

Figs.\ \ref{fig:06}a and \ref{fig:06}b show the variation of the two-phase friction multiplier with the superficial gas and liquid velocities in hydrophilic and hydrophobic PMF flow fields, respectively. It can be seen that at the same air velocity, the two-phase friction multiplier increases with the increase of the superficial liquid velocity, especially at low air velocity, which indicates that the liquid retention in the PMF flow field increases. In addition, at low air velocity, the retention of liquid water in a large area leads to the two-phase friction multiplier significantly greater than unity. Therefore, the film flow pattern and the slug flow pattern show the highest two-phase flow resistance. With the increase of airflow rate, the two-phase friction multiplier decreases and gradually approaches unity, which indicates that the pressure drop during the ligament and droplet flow pattern is almost the same as the value in the single-phase gas flow. This is because the droplets and thin films trapped on the ligament wall have a minor effect on the airflow in the pore center, which is consistent with the visualization images.

\subsubsection{Transient pressure drop analysis in the time domain}\label{sec322}
\begin{figure}
  \centering
  \includegraphics[scale=0.7]{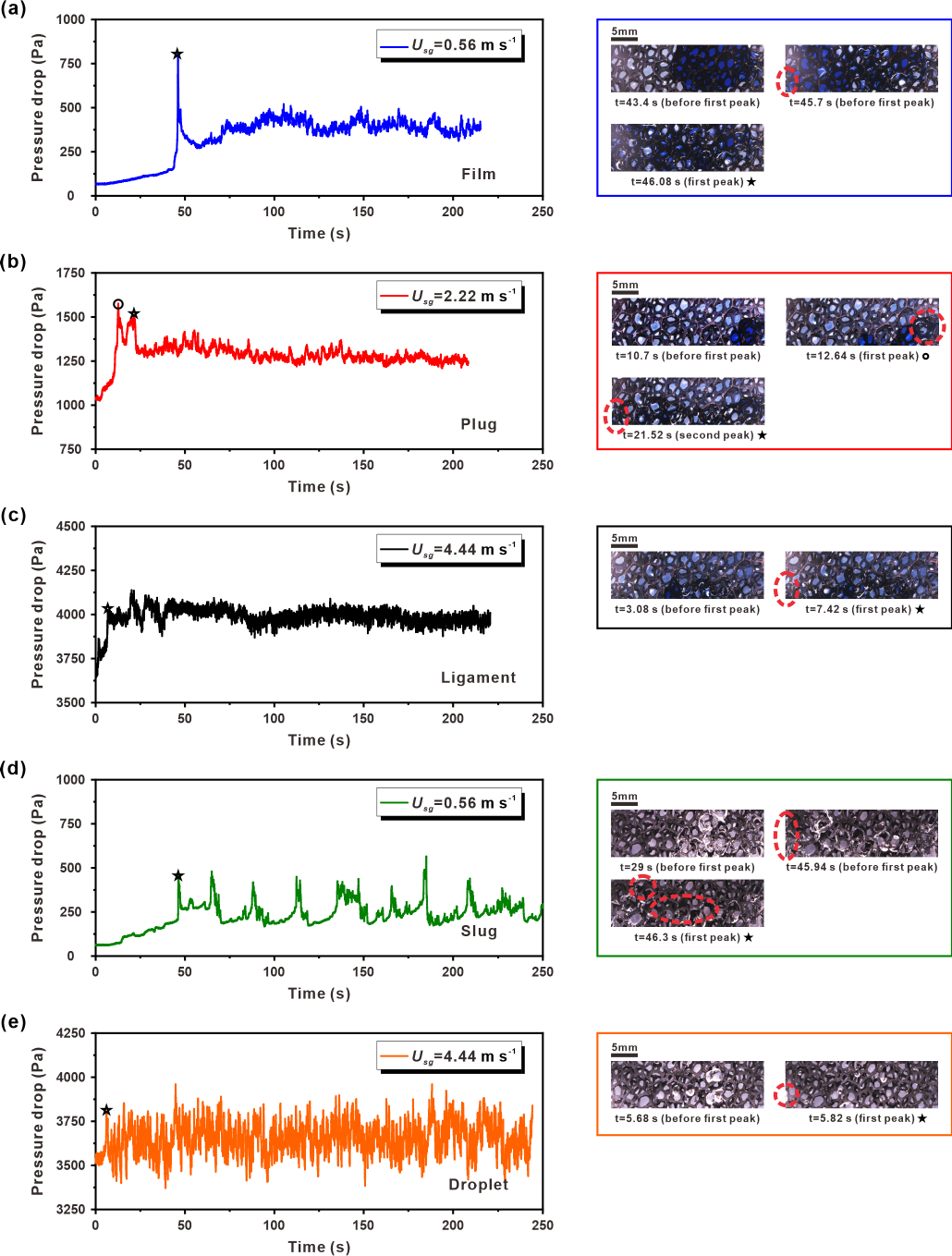}
  \caption{The left column shows typical curves of pressure drop for five flow patterns in the PMF flow field. The stars mark the first instant when the liquid water reaches the outlet edge of the foam and is partially discharged, and the circle marks the moment when the gas plug begins to form in the continuous liquid film. The right column shows the experimental images corresponding to the flow patterns, and the images corresponding to the peaks of the pressure drop curves are selected. The red dotted circles mark the position of the movement of liquid water. (a) Film flow; (b) Plug flow; (c) Ligament flow; (d) Slug flow; (e) Droplet flow. The superficial liquid velocities are constant at 0.01068 m s$^{-1}$ and the superficial gas velocities are (a) 0.56 m s$^{-1}$; (b) 2.22 m s$^{-1}$; (c) 4.44 m s$^{-1}$; (d) 0.56 m s$^{-1}$; (e) 4.44 m s$^{-1}$, respectively. Figs.~\ref{fig:07}(a-c) were measured in the hydrophilic PMF flow field, and Figs.~\ref{fig:07}(d-e) were measured in the hydrophobic PMF flow field.}\label{fig:07}
\end{figure}

The hydrodynamics in porous media can be described by the relationship of the pressure drop with the geometry, the fluid properties, and the flow parameters of porous media \cite{Tao2010MetallicFoam3DFlowSimulations, Tao2015PorousMediaSimulation}, which is still unclear for the two-phase flow in the PMF flow field but very important for the PEMFC applications. In this study, the pressure drop between the inlet and outlet of the PMF flow field was measured in real time by a differential pressure transducer. Figs.\ \ref{fig:07}(a-e) show the typical pressure fluctuation signal (PFS) curves for different flow patterns. The initial point of all PFS curve measurements corresponds to the single gas phase pressure drop at this air velocity.

The PFS curve of the film flow shows a single peak, as marked by the star in Fig.\ \ref{fig:07}a, which represents the first instant when the liquid water reaches the outlet edge of the foam ($t = 45.7$ s), and the discharge of part of the water at the outlet ($t = 46.08$ s) causes the instantaneous fluctuation of the pressure drop. During the film flow test, a stable water path along the airflow direction is established based on the strong capillarity. Subsequent water tends to flow along this path till the outlet. Fig.\ \ref{fig:07}b shows that the plug flow has two peaks of the PFS curve marked by a circle and a star, corresponding to the formation of the gas plug in the continuous liquid film ($t = 12.64$ s) and the partial discharge of liquid water at the edge of the foam outlet ($t =21.52$ s) for the first time. As shown in Fig.\ \ref{fig:07}c, with increasing the airflow velocity, the amplitude of pressure drop fluctuation is weakened, indicating the transition to the ligament flow pattern. In the ligament flow, water adheres to the hydrophilic metal foam ligament surface as a flat thin film, which results in lower fluctuation of the pressure drop in this flow pattern. The PFS curve in the slug flow has multiple peaks periodically, as shown in Fig.\ \ref{fig:07}d. Each peak indicates the movement or removal of a certain amount of liquid water ($t = 46.3$ s). The climbing stage between two peaks in the PFS curve corresponds to the slow accumulation of the slug. The smaller the air velocity is, the longer the process lasts. Further increase of air velocity results in a larger pressure drop in the droplet flow pattern, as shown in Fig.\ \ref{fig:07}e. The pressure drop fluctuation is a consequence of the random dispersion and movement of droplets in the PMF flow field.

\subsubsection{Transient pressure drop analysis in the frequency domain}\label{sec323}
\begin{figure}
  \centering
  \includegraphics[scale=0.7]{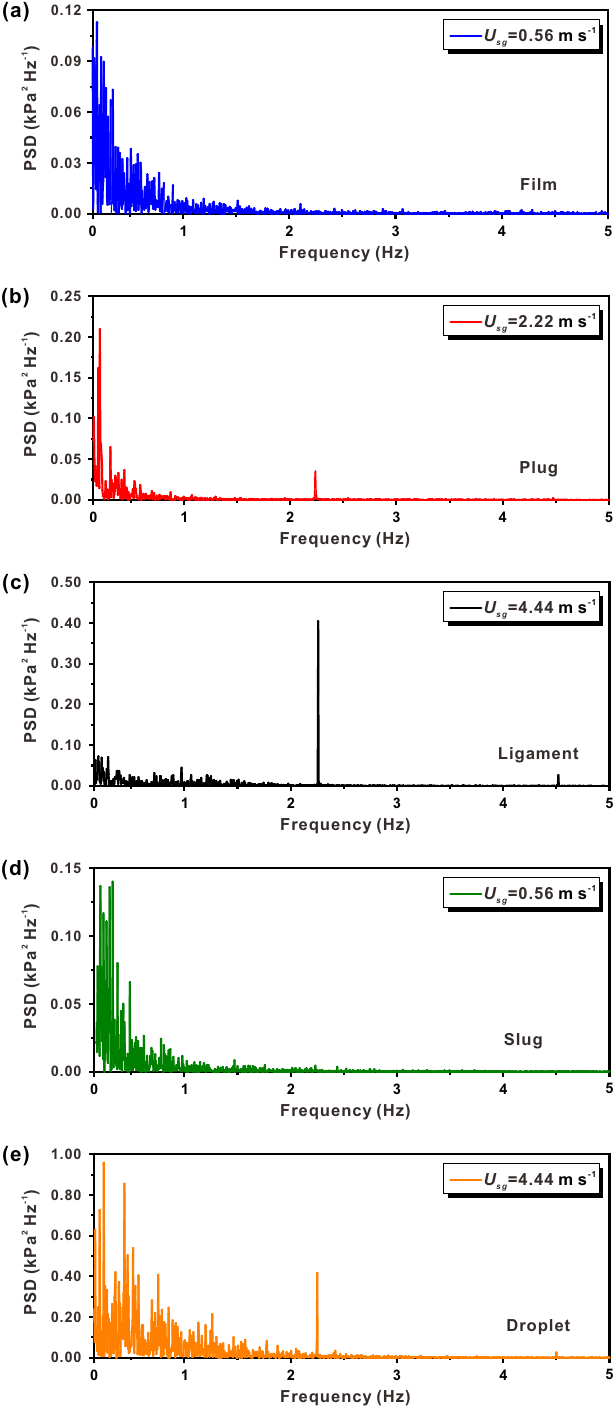}
  \caption{Normalized power spectral density function for pressure drop signals during different flow patterns for two-phase flow in the PMF flow field: (a) Film flow; (b) Plug flow; (c) Ligament flow; (d) Slug flow; (e) Droplet flow. The superficial liquid velocities are constant at 0.01068 m s$^{-1}$ and the superficial gas velocities are (a) 0.56 m s$^{-1}$; (b) 2.22 m s$^{-1}$; (c) 4.44 m s$^{-1}$; (d) 0.56 m s$^{-1}$; (e) 4.44 m s$^{-1}$, respectively. Figs.~\ref{fig:08}(a-c) were measured in the hydrophilic PMF flow field, and Figs.~\ref{fig:08}(d-e) were measured in the hydrophobic PMF flow field.}\label{fig:08}
\end{figure}

To analyze the above pressure drop fluctuation and the frequency characteristics corresponding to different flow patterns, we calculated the power spectral density (PSD) of the pressure drop fluctuation, which is the absolute value of the squared norm of the Fourier transform of time series data. The time series data here refers to the instantaneous pressure drop from the inlet to the outlet of the PMF flow field, following the approach presented in Ref.~\cite{Tao2020PressureSignalAnalysis}, as shown in Fig.\ \ref{fig:08}. By signal analysis, we can reduce the subjectivity of visualization in the process of identifying and characterizing different flow patterns \cite{Tao2020PressureSignalAnalysis}.

The PSD of the pressure drop indicates that different flow patterns have different characteristic frequency bands, representing the two-phase flow characteristics, as shown in Fig.\ \ref{fig:08}. For film flow, a bandwidth of peak frequency ranging from 0 to 0.9 Hz is seen due to many pores being filled with liquid water, as shown in Fig.\ \ref{fig:08}a. The plug flow pattern, as shown in Fig.\ \ref{fig:08}b, has the dominant spectrum of a low-frequency band and a sharp peak at 0--0.7 Hz, which is related to the air plug formed by the growth and rupture of minute bubbles in the continuous liquid films. It is worth noting that the plug flow and the film flow have similar peak frequencies in the PSD curves, which is attributed to the spreading of the liquid film in the local region, but the bandwidth for film flow is wider than that for plug flow. Fig.\ \ref{fig:08}c shows the PSD of the ligament flow has obvious characteristics of the high-frequency dominant spectrum, and the intensity of the low-frequency stage is significantly lower than that of the film flow and the plug flow, because, under the effect of high-speed airflow and strong adhesion of hydrophilic ligament, liquid water is mainly retained on the ligament surface. There is no large area of liquid film spreading in the pore center. A high-frequency peak of 2.25 Hz is observed which is the result of high-speed airflow in the pores. Meanwhile, the powerful energy of the peak increases with the increase of airflow rate, indicating that the flow pressure drop becomes higher.

The periodic accumulation and breakup of slugs results in broadband of dominant frequency in the slug flow pattern, about 0--1 Hz, as shown in Fig.\ \ref{fig:08}d. In the slug flow pattern, liquid water would fill up several pores, and the liquid holdup in the PMF flow field is high, and the airflow is greatly hindered. Therefore, the dominant high-frequency peak, which corresponds to the air-phase flow in the pore center, is absent. The breakup and coalescence of the liquid droplets are related to several different frequencies, so the droplet flow pattern has multiple wider dominant frequency bands, as shown in Fig.\ \ref{fig:08}e. The same dominant frequency band centered at 2.25 Hz also exists in the droplet flow pattern, indicating that the droplets are mainly located at the intersection of the ligaments and do not cause large pressure drop fluctuations to the airflow in the pore center. This feature in the PSD curve is consistent with the images taken in the experiment, indicating that the hindrance of liquid water to airflow is weakened in the droplet flow pattern.

From the above features of the frequency information of the pressure drop, we can see that the hydrodynamics under different flow patterns determine the dominant frequency corresponding to each flow pattern.  In the hydrophilic PMF flow field, the transition of the flow pattern from film flow to plug flow can be judged by mainly observing the bandwidth of the dominant frequency band of the PSD curve. The transition of the flow pattern from plug flow to ligament flow can be judged mainly by observing whether there is a high-frequency dominant frequency band centered at 2.25 Hz and its intensity in the PSD curve. Moreover, the intensity of the low-frequency stage in ligament flow is significantly lower than that of plug flow. In the hydrophobic PMF flow field, judging the transition of slug flow to droplet flow is to observe whether the PSD curve has a high-frequency dominant frequency band centered at 2.25 Hz, and the peak intensity of droplet flow is much higher than that of slug flow. Thus, the change of the dominant frequency can be used as an indicator to analyze the transition between different flow patterns when direct flow visualization is impossible.

\section{Conclusions}\label{sec4}
In summary, we have studied the gas-liquid two-phase flow in the PMF flow field via direct optical visualization and investigated the effects of foam pore size, surface wettability, and gas and liquid superficial velocities on the two-phase flow pattern under the working conditions of the PEMFCs. Five different flow patterns are observed in the PMF flow field, namely film flow, plug flow, ligament flow, slug flow, and droplet flow. For the PMF with small pores, the adhesion and spreading effect of the liquid film is stronger because the capillarity is dominant compared with the aerodynamic forces of the airflow. Hydrophilic foam has greater ligament surface adhesion force and air transport resistance, resulting in severe liquid retention. Due to the decrease of ligament adhesion and the increase of air aerodynamic force, the air transport is strengthened and the water coverage is significantly reduced in the hydrophobic PMF flow field. These results suggest that it is necessary to carry out hydrophobicity treatment to the PMF flow field from the point of strengthening mass transfer and preventing water flooding for practical applications.

We have also analyzed pressure drop fluctuation, which is critical to quantify water accumulation and characterize different flow patterns and their transitions. The flow resistance analysis shows that the film flow and slug flow show the highest two-phase friction multiplier because of the accumulation of water in a large area in the PMF flow field. The two-phase friction multiplier during the ligament flow and the droplet flow pattern is the smallest and approaches unity with the increase of airflow rate. The film flow and the plug flow have the significant characteristics of the dominant spectrum with the low-frequency bandwidth, while the slug flow has a wide dominant frequency bandwidth. The breakup and merging of droplets are related to several different contribution frequencies. The frequency analysis also shows that the dominant frequency can be used as an indicator to analyze the transition between different flow patterns when direct flow visualization is impossible. This could be an important tool for flow diagnostics and online monitoring in both laboratory experiments and industrial application processes in PEMFCs as direct flow visualization is always a challenge due to the material properties used in PEMFCs, particularly for in-situ PEMFC measurements. The ex-situ flow visualization also provides a way to decouple electrochemical reaction and heat/mass transfer from complex flow phenomena inside PEMFCs.

\section*{Declaration of Competing Interest}
The authors declare that they have no known competing financial interests or personal relationships that could have appeared to influence the work reported in this paper.

\section*{Acknowledgements}
This work is supported by the National Natural Science Foundation of China (Grant Nos.\ 51920105010, 52176083, and 51921004), and the Department of Science and Technology of Inner Mongolia (Grant No.~2022JBGS0027).

\bibliography{PorousFlowField}

\end{document}